\documentclass[%
aps,%
showpacs,%
preprintnumbers,
floatfix,%
superscriptaddress,%
nofootinbib%
]{revtex4}

\usepackage{graphicx,amsmath,amssymb}

\begin{document}


\preprint{OU-HET-688/2010 }

\title{
  Unified description of quark and lepton mixing matrices 
  based on a Yukawaon model

}

\author{Hiroyuki Nishiura} %
\email{nishiura@is.oit.ac.jp} %
\affiliation{%
Faculty of Information Science and Technology, 
Osaka Institute of Technology, 
Hirakata, Osaka 573-0196, Japan
}
\author{Yoshio~Koide}
\email{koide@het.phys.sci.osaka-u.ac.jp} %
\affiliation{%
  Department of Physics, Osaka University,  
Toyonaka, Osaka 560-0043, Japan
}

\date{February 10, 2011}

\begin{abstract}
Based on a supersymmetric Yukawaon model with O(3) family symmetry, 
possible forms of quark and lepton mixing matrices are systematically 
investigated under a condition that the up-quark mass matrix form leads to 
the observed nearly tribimaximal mixing in the lepton sector.
Although the previous model could not provide a good fitting of 
the observed quark mixing, the present model can give a reasonably good fitting 
not for lepton mixing but also the quark mixing 
by using a different origin of the $CP$ violation from one in the previous model.

\end{abstract}

\pacs{
  12.60.-i, 
  11.30.Hv, 
}

\maketitle

\section{Introduction}

One of the current subjects of the particle flavor physics is 
to understand quark and lepton masses and mixings.
The investigation of them, even if it is phenomenological, will 
provide a promising clue to new physics.
The so-called ``tribimaximal" mixing observed in the neutrino
mixing \cite{atm,solar} is very suggestive of a fundamental law in the lepton
sector.
Usually, the observed tribimaximal mixing has been explained
by assuming a discrete symmetry \cite{tribi}.

Meanwhile, as a neutrino mass matrix model without assuming
any discrete symmetry, an unfamiliar model 
\cite{MNS_yukawaon08,O3PLB09} has been proposed by using 
a seesaw type neutrino mass matrix $M_\nu = m_D M_R^{-1} m_D^T$.
In this model, the Dirac neutrino mass matrix $m_D$ is given by 
$m_D \propto M_e$ ($M_e$ is a charged lepton mass matrix) and the Majorana
mass matrix $M_R$ of the right-handed neutrinos is given by
$$
M_R \propto M_u^{1/2} M_e +M_e M_u^{1/2} .
\eqno(1.1)
$$
Here $M_u^{1/2}$ is defined by $M_u^{1/2} (M_u^{1/2})^T =M_u$
($M_u$ is an up-quark mass matrix with a symmetric form
$M_u^T =M_u$).
The model \cite{O3PLB09} can lead to a nearly tribimaximal 
mixing by assuming suitable up-quark mass matrix 
as we give a short review in the next section.
The model has only four parameters: one ($\xi_\nu$) is in the neutrino 
sector, and one ($a_u$) is in the up-quark sector, and two 
($a_d e^{i\alpha_d}$) in the down-quark sector.  
(Here, we consider that the charged lepton mass  
$(m_e, m_\mu, m_\tau)$ are known parameters, and we do not
count these parameters as adjustable parameters.)
This model leads to excellent fitting for up-quark mass ratios
$m_u/m_c$ and $m_c/m_t$ and neutrino mixing parameters
$\sin^2 2\theta_{atm}$, $\tan^2 \theta_{solar}$ and $|U_{e3}|^2$,
only by adjusting the two parameters $a_u$ and $\xi_\nu$.  
On the other hand, for down-quark sector, the fitting is not so excellent,
especially, the predicted values of $|V_{ub}|$ and $|V_{td}|$
are somewhat large compared with the observed values as far
as we use parameter values which can give reasonable values 
for the observed down-quark mass ratios. 

The purpose of the present paper is to investigate an improved 
version of the above model and  to search systematically for 
parameter values which can give reasonable quark mass ratios,  
quark mixing parameters (Cabibbo-Kobayashi-Maskawa (CKM) 
mixing matrix) and neutrino mixing parameters. 
In Sec.II, we will show that the four parameter model cannot have
reasonable parameter region consistent with four quark mass ratios, 
three neutrino mixing parameters, and four CKM mixing parameters.
In Sec.III, we propose a revised model and give parameter fitting 
for 11 observables.
(In the present model, we do not discuss the observed value
$R \equiv \Delta m^2_{solar}/\Delta m^2_{atm}$ for neutrino masses, because 
we can always have an additional one parameter which inevitably 
appears in the model and affects only the mass ratios
$R$, but does not affect neutrino mixing and observables in the
 quark sector.)   
Finally, Sec.IV is devoted to the summary and discussions.


\section{Supersymmetric Yukawaon model}

In this section, we give a short review of a quark and lepton 
mass matrix model \cite{O3PLB09} based on the supersymmetric 
Yukawaon model, because, in this paper, we propose a revised 
version of this model.

In the Yukawaon model, we put the following assumption:

\noindent (i) We consider that the Yukawa coupling constants 
are effectively given by
$$
Y_f^{eff} = y_f \frac{\langle Y_f \rangle}{\Lambda} ,
\eqno(2.1)
$$
where $\langle Y_f \rangle$ ($f=u,d,e,\nu$ and so on) 
are vacuum expectation values (VEVs) 
of new scalars $Y_f$ with $3\times 3$ components of 
O(3) family symmetry and 
$\Lambda$ is an energy scale of the effective theory. 
(For the time being, we assume $\Lambda \sim 10^{14-15}$ GeV.) 
Therefore, the would-be Yukawa interactions are given by
$$
H_{Y}= \sum_{i,j} \frac{y_u}{\Lambda} u^c_i(Y_u)_{ij} {q}_{j} H_u  
+\sum_{i,j}\frac{y_d}{\Lambda} d^c_i(Y_d)_{ij} {q}_{j} H_d 
$$
$$
+\sum_{i,j} \frac{y_\nu}{\Lambda} \ell_i(Y_\nu)_{ij} \nu^c_{j} H_u  
+\sum_{i,j}\frac{y_e}{\Lambda} \ell_i(Y_e)_{ij} e^c_j H_d +{\rm H.c.} 
+ \sum_{i,j}y_R \nu^c_i (Y_R)_{ij} \nu^c_j ,
\eqno(2.2)
$$
where $q$ and $\ell$ are SU(2)$_L$ doublet fields, and
$f^c$ ($f=u,d,e,\nu$) are SU(2)$_L$ singlet fields.

\noindent (ii) In order to distinguish each $Y_f$ from others, 
we assume a U(1)$_X$ symmetry (i.e. ``sector charge")
in addition to the O(3) symmetry, and we have assigned 
U(1)$_X$ charges as $Q_X(Y_f)=x_f$, $Q_X(f^c)=-x_f$ and
$Q_X(Y_R)=2 x_\nu$.
(The SU(2)$_L$ doublet fields $q$, $\ell$, $H_u$ and $H_d$
are assigned to sector charges $Q_X=0$.) 

\noindent (iii) For the neutrino sector, we assume 
$Q_X(\nu^c) =Q_X(e^c)$, so that the Yukawaon $Y_e$ can also 
couple to the neutrino sector as $(\ell Y_e \nu^c) H_u$ 
instead of $(\ell Y_\nu \nu^c) H_u$ in Eq.(2.2).
Therefore, we can change the above model into a model
without $Y_\nu$.
Hereafter, we read $Y_\nu$ in Eq.(2.2) as $Y_e$.
Besides, we can have a term
$$
 \sum_{i,j,k} \frac{y'_R}{\Lambda}\nu^c_i (Y_e)_{ik} (Y_e)_{kj}\nu^c_j ,
\eqno(2.3)
$$
in addition to the right-hand side of Eq.(2.2),
because $Y_eY_e$ has the same U(1)$_X$ charge as $Y_R$,
i.e. $Q_X=2 x_e$. 
Although this term (2.3) leads to an additional neutrino 
mass term, the term does not affect neutrino mixing \cite{inverseM_nu} as far as the neutrino mass matrix $M_\nu$
is real, \footnote{
When $R^T M_1 R=D_1 \equiv {\rm diagonal}$, 
the inverse matrix $M_1^{-1}$ is 
also diagonalized as $R^T M_1^{-1} R= D_1^{-1}$ by
the same orthogonal transformation matrix $R$; 
When we take $M=M_1+m_0 {\bf 1}$, $M$ is diagonalized 
as $R^T M R= D_1 +m_0 {\bf 1}$, so that we can 
diagonalize $M^{-1}$ as $R^T M^{-1} R = (R^T M R)^{-1}
=(D+m_0{\bf 1})^{-1}$. 
}
because of $M_\nu \propto Y_e [(\cdots) + Y_e Y_e]^{-1} Y_e =
[Y_e^{-1} (\cdots) Y_e^{-1} + {\bf 1}]^{-1}$. 

\noindent
(iv) We give a superpotential $W$ which is invariant under 
O(3) family symmetry and U(1)$_X$ symmetry, and solve 
supersymmetric (SUSY) vacuum conditions.
As a result, we obtain VEV relations among 
Yukawaons.

For example, we have assumed the following superpotential
$$
W_e = \lambda_e {\rm Tr}[\Phi_e \Phi_e \Theta_e] 
+\mu_e {\rm Tr}[Y_e \Theta_e] + W_\Phi. 
\eqno(2.4)
$$
Here we have assumed $Q_X(\Phi_e) = \frac{1}{2}Q_X(Y_e) =
-\frac{1}{2}Q_X(\Theta_e)$ and the term $W_\Phi$ has been
introduced in order to determine a VEV spectrum 
$\langle \Phi_e \rangle$ completely. 
Then, from a SUSY vacuum condition
$$
\frac{\partial W}{\partial \Theta_e} = \lambda_e \Phi_e \Phi_e +
\mu_e Y_e =0 ,
\eqno(2.5) 
$$
we obtain a VEV relation
$$
\langle Y_e \rangle =
-\frac{\lambda_e}{\mu_e} \langle \Phi_e\rangle  \langle \Phi_e\rangle . 
\eqno(2.6)
$$ 
The VEV value $\langle \Phi_e\rangle$ is 
derived from the term $W_\Phi$ (for example, see
Refs.\cite{Sumino09PLB,Sumino09JHEP,e-spec_10PLB}).  
However, for simplicity, in this paper, we use the observed 
values of the charged lepton masses
straightforwardly for the VEV value as given by
$$
\langle \Phi_e \rangle_e = {\rm diag}(v_1,v_2,v_3) 
\propto {\rm diag}(\sqrt{m_e}, \sqrt{m_\mu}, \sqrt{m_\tau}).
\eqno(2.7)
$$
In other words, we have assumed the ad hoc relation (2.7), 
the derivation of which is not discussed in the present paper.
Hereafter, for counting a number of ``adjustable" 
parameters, we do not include $v_i$ in the number.
Here, the notation $\langle A \rangle_f$ denotes a form of 
a VEV matrix $\langle A \rangle$ in the diagonal basis of 
$\langle Y_f\rangle$ (we refer to it as $f$ basis).
The scalar $\Theta_e$ does not have a VEV, i.e. 
$\langle \Theta_e \rangle =0$.
Therefore, terms which include more than two of $\Theta_e$
do not play any physical role, so that we do not consider 
such terms in the present effective theory. 
[Hereafter, we will denote fields whose VEV values are zero
as notations $\Theta_A$ ($A=u, d, \cdots$).]

Next, for the purpose of the comparison of our new model with the previous one,  
we give a short review of quark and lepton mass matrix forms of the previous model discussed 
in Ref.\cite{O3PLB09}. 
The explicit form of the superpotential for the previous model is given in 
Ref.\cite{O3PLB09}. That, for the new model, shall be given in the next section.

In the previous model\cite{O3PLB09}, the quark mass matrices, i.e. $\langle Y_u \rangle$ and 
$\langle Y_d \rangle$, are given  as
$$
M_u \propto \langle Y_u \rangle \propto \langle \Phi_u \rangle
\langle \Phi_u \rangle , 
\eqno(2.8)
$$
$$
\langle \Phi_u \rangle_e \propto \langle \Phi_e \rangle_e 
\left(  \langle E \rangle_e + a_u \langle  X \rangle_e 
\right) \langle \Phi_e \rangle _e ,
\eqno(2.9)
$$ 
$$
M_d \propto 
\langle Y_d \rangle_e \propto \langle \Phi_e \rangle_e 
\left(  \langle E \rangle_e +
a_d e^{i\alpha_d} \langle X \rangle_e 
\right) \langle \Phi_e \rangle _e ,
\eqno(2.10)
$$
respectively. 
(For convenience, we have changed the definitions of
$a_u$ and $a_d$ from those in Ref.\cite{O3PLB09}.)
Here, $\langle X \rangle_e$ and 
$\langle E \rangle_e$ are given by
$$
\langle X \rangle_e = X v_X \equiv \frac{1}{3} \left(
\begin{array}{ccc}
1 & 1 & 1 \\
1 & 1 & 1 \\
1 & 1 & 1 
\end{array} \right) v_X, \ \ \ \ \ 
\langle E \rangle_e = {\bf 1} v_E \equiv  \left(
\begin{array}{ccc}
1 & 0 & 0 \\
0 & 1 & 0 \\
0 & 0 & 1 
\end{array} \right) v_E .
\eqno(2.11)
$$
(Here, the VEV form $\langle X \rangle_e$ 
breaks the O(3) flavor symmetry into S$_3$.)
Therefore, we obtain quark mass matrices
$$
M_u^{1/2} \propto  M_e^{1/2} \left(   {\bf 1} 
+ a_u {X} \right) M_e^{1/2} , \ \ \ 
M_d \propto  M_e^{1/2} \left( {\bf 1} +a_d e^{i\alpha_d}  {X} 
 \right) M_e^{1/2} ,
\eqno(2.12)
$$ 
on the $e$ basis.  
Here, we have redefined the coefficients $a_u v_X/v_E$ and 
$a_d v_X/v_E$ in Eqs.(2.9) and (2.10) as $a_u$ and $a_d$,
respectively. 
Hereafter, for numerical estimate of $a_u$ and $a_d$, we 
use the definition of those in Eq.(2.12).
(This quark mass matrix form (2.12) has first been proposed 
in Ref.\cite{DemocSeesaw} as a ``democratic universal seesaw
mass matrix model".)  
Note that we have assumed that the O(3) relations are valid
only on the $e$ and $u$ bases, so that $\langle Y_e\rangle$ 
and $\langle Y_u\rangle$ must be real. 
[The VEV matrix $\langle\Phi_u\rangle$ must satisfy the 
relation (2.9) on the $e$ basis, while $\langle\Phi_u\rangle$ 
must also satisfy the relation 
$\langle Y_u\rangle \propto \langle\Phi_u\rangle_u 
\langle\Phi_u\rangle_u$ on the $u$ basis.
However, for the down-quark sector, such a condition is
not required, because  $\langle Y_d\rangle$ is given by
Eq.(2.10) only on the $e$ basis.]

A case $a_u \simeq -1.79$ can give a reasonable up-quark
mass ratios $\sqrt{{m_{u1}}/{m_{u2}}}=0.043$ and   
$\sqrt{{m_{u2}}/{m_{u3}}}=0.057$, which are in favor of
the observed values \cite{q-mass}
$$
\sqrt{\frac{m_{u}}{m_{c}}}=0.045^{+0.013}_{-0.010} ,
\ \ \ \ 
\sqrt{\frac{m_{c}}{m_{t}}}=0.060\pm 0.005 ,
\eqno(2.13)
$$ 
at $\mu=m_Z$. 

In this paper, we will carry out 
parameter-fitting at $\mu=m_Z$, because we interest
in the mixing values at $\mu=m_Z$.
Exactly speaking, fitting for the mass ratios must
be done at $\mu=\Lambda \sim 10^{14}$ GeV.
However, at present, our model does not intend to give 
so precise predictions of the quark mass ratios.
For example, we know \cite{q-mass} 
$\sqrt{m_u/m_c}=0.046^{+0.013}_{-0.012}$ and
$\sqrt{m_c/m_t}=0.051^{+0.002}_{-0.006}$ even at
$\mu=2\times 10^{16}$ GeV ($\tan\beta=10$).
Even in $\sqrt{m_c/m_t}$, the discrepancy is smaller
than 20\%. 
Besides, the mass values are dependent on the value of  
$\tan\beta$ in the SUSY model.
Therefore, for simplicity, in this paper, we will carry 
out the parameter-fitting at $\mu=m_Z$.

On the other hand, in the neutrino mass matrix 
$$
M_\nu \propto \langle Y_e \rangle_e \langle Y_R \rangle_e^{-1}
\langle Y_e \rangle_e ,
\eqno(2.14)
$$
the Majorana mass matrix of the right-handed neutrinos
$\langle Y_R \rangle_e$ is given by
$$
\langle Y_R \rangle \propto 
\langle Y_e \rangle_e \langle P_u \rangle_e 
\langle\Phi_u\rangle_e + \langle\Phi_u\rangle_e 
\langle P_u \rangle_e \langle Y_e \rangle_e 
+\xi_\nu (\langle P_u \rangle_e \langle Y_e \rangle_e \langle\Phi_u\rangle_e
 + \langle\Phi_u\rangle_e \langle Y_e \rangle_e \langle P_u \rangle_e) .
\eqno(2.15)
$$
Here, we have introduced a new field $P_u$ with a VEV
$$
\langle P_u \rangle_u \propto {\rm diag}(+1,-1,+1) ,
\eqno(2.16)
$$
in order to make ``effective" eigenvalues of 
$\langle \Phi_u\rangle_u$ positive, 
because the eigenvalues of
$\langle \Phi_u\rangle_u=(v_{u1}, v_{u2}, v_{u3})$ give signs 
$(+,-,+)$ for the parameter value $a_u \sim -1.8$.
(The field $P_u$ has been introduced from a phenomenological
reason.  If the factor (2.16) were absence [i.e. $Y_R$ 
were given by $\langle Y_R \rangle \propto 
\langle Y_e \rangle_e \langle\Phi_u\rangle_e + 
\langle\Phi_u\rangle_e \langle Y_e \rangle_e$], 
we could not give the observed maximal 
mixing $\sin^2 2\theta_{atm} \simeq 1$ \cite{atm} for 
any values of the parameters $a_u$ and $\xi_\nu$.)
The reason for the $\xi_\nu$ term in Eq.(2.15) is as follows:
When we consider a term $Y_e P_u \Phi_u + \Phi_u P_u Y_e$ 
we must also consider an existence of 
$P_u Y_e \Phi_u+\Phi_u Y_e P_u$ \cite{O3PLB09}, 
because they have the same U(1)$_X$ charges.
The results at $a_u \simeq -1.78$ are excellently in favor
of the observed neutrino oscillation parameters
$\sin^2 2\theta_{atm}=1.00_{-0.13}$ \cite{atm} and
$\tan^2 \theta_{solar}=0.469^{+0.047}_{-0.041}$ \cite{solar} 
by taking a small value of $|\xi_\nu|$, $\xi_\nu=+0.005$ or
 $\xi_\nu=-0.0012$.

Thus, the model in Ref.\cite{O3PLB09} can successfully 
fit two up-quark mass ratios and three neutrino mixing parameters 
only by the two parameters $a_u$ and $\xi_\nu$.
On the other hand, the fitting of six observable quantities 
(two down-quark mass ratios and four CKM mixing parameters) only by
two parameters $a_d$ and $\alpha_d$ given in Ref.\cite{O3PLB09} are not in
excellent agreement with the observed values.
Especially, the predicted values of $|V_{ub}|$ and $|V_{td}|$
are considerably larger than the observed ones.
We find from a systematical parameter search that 
this is not due to incompleteness of the parameter search, 
but plausible values of the CKM mixing  
parameters cannot be obtained even if we abandon the fitting of the down-quark 
mass ratios. 

Considering the success in the up-quark and neutrino sectors,
we do not change the model for the up-quark and neutrino sectors.
We fix the parameter values as $a_u \sim -1.8$.
The observed values of the down-quark masses are 
as follows \cite{q-mass}
$$
{\frac{m_{d}}{m_{s}}}=0.053^{+0.051}_{-0.029} ,
\ \ \ \ 
{\frac{m_{s}}{m_{b}}}=0.019\pm 0.006 ,
\eqno(2.17)
$$ 
at $\mu=m_Z$. 
We consider that the mass ratio $m_d/m_s$ may be sensitive 
due to an unknown effect of a minor change of the model, 
so that, for the time being, we disregard the fitting of 
$m_d/m_s$ and concentrate on the fitting of $m_s/m_b$. 
Although a parameter value $a_d \sim -16$ can give 
a reasonable prediction of $m_s/m_b$, the solution cannot give
reasonable predictions of the CKM mixing parameters, we
rule out this solution $a_d \sim -16$.
We find that there are another solutions of $a_d$ in a range 
$-a_d = 1.3 -1.7$,
which can roughly give $m_s/m_b = 0.1-0.4$. 
The solutions have a possibility that they can
give reasonable values of the CKM mixing parameters. 
Therefore, we investigate the case with $-a_d = 1.3 -1.7$
in detail.

\begin{figure}[t!]
  \includegraphics[width=78mm,clip]{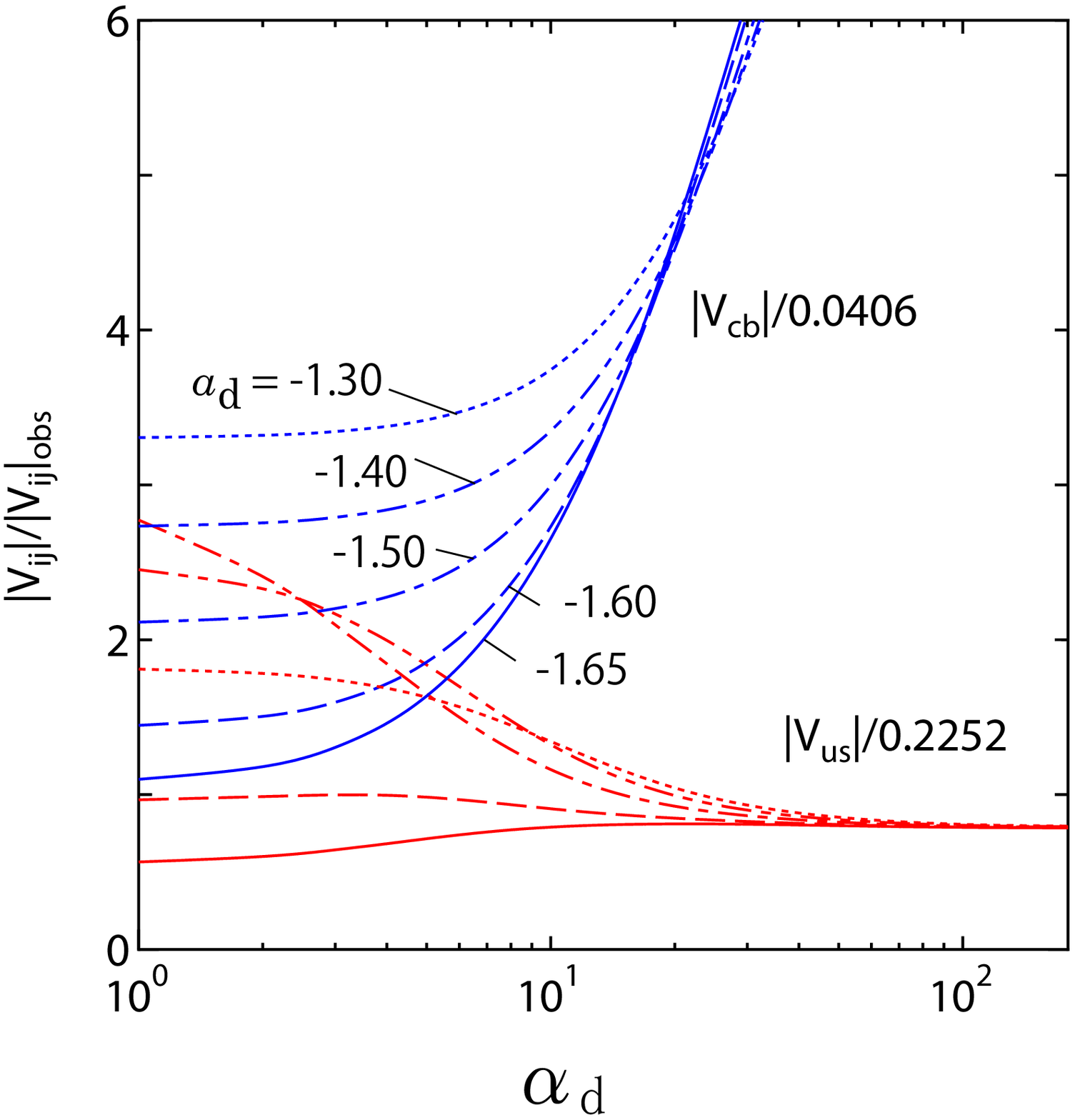}
  \includegraphics[width=78mm,clip]{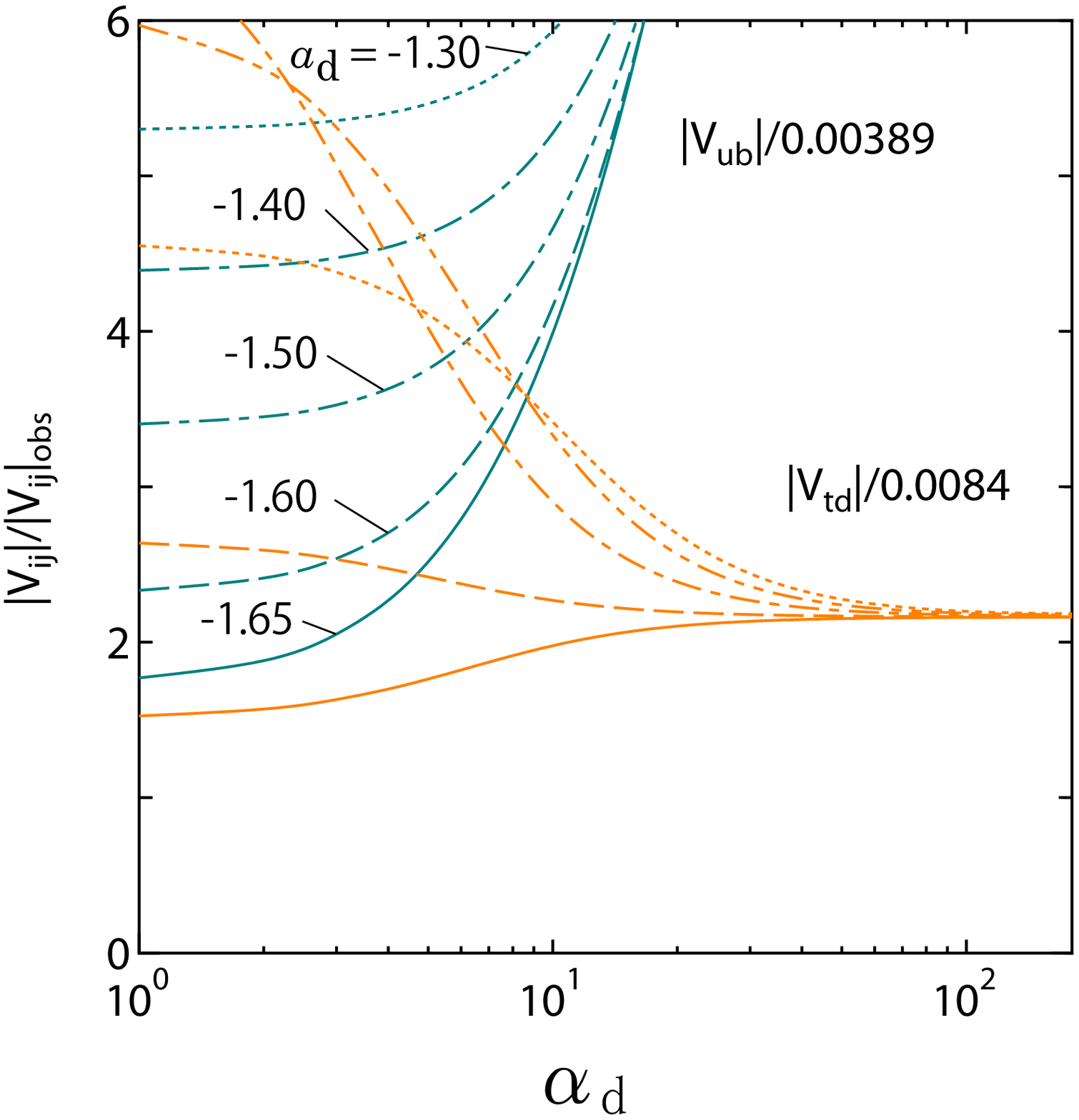}  
  
  \caption{$|V_{ij}|/|V_{ij}|_{obs}$ versus a phase parameter
  $\alpha_d$. 
(a) $|V_{us}|/0.2252$ and $|V_{cb}|/0.0406$; 
(b) $|V_{ub}|/0.00389$ and $|V_{td}|/0.0084$.
The parameter $a_u$ in the up-quark sector is
fixed at $a_u=-1.79$. 
The five curves represent $a_d=-1.30$, $-1.40$, $-1.50$,
$-1.60$ and $-1.65$, respectively.
}
  \label{old_Vij}
\end{figure}

The results are shown in Fig.1, where the predicted values
$|V_{ij}|$ versus the phase parameter $\alpha_d$ are given 
in the unit of the observed values \cite{PDG10} 
$|V_{ij}|_{obs}$ 
$$
\begin{array}{ll}
|V_{us}|_{obs}=0.2252\pm 0.0009, & 
|V_{cb}|_{obs}=0.0406\pm 0.0013, \\
|V_{ub}|_{obs}=0.00389 \pm 0.00044 , &
|V_{td}|_{obs}=0.0084 \pm 0.0006 .
\end{array}
\eqno(2.18)
$$
Here, we have illustrated the behaviors of $|V_{ij}|$
for the range $\alpha_d=0^\circ - 180^\circ$, because 
the behaviors for $\alpha_d=360^\circ - 180^\circ$ are 
just the same as that for $\alpha_d=0^\circ - 180^\circ$.
As seen in Fig.1(a), in order to obtain a reasonable 
value of $|V_{cb}|$, we must choose a value
of $\alpha_d$ smaller than $\alpha_d \sim 10^\circ$, and 
also a value of $a_d$ smaller than $-a_d \sim 1.5$.
However, from Fig.1(b), we can conclude there is no 
solution for a reasonable value of $|V_{us}|$ for 
any values of $a_d$ and $\alpha_d$ even at the cost
of the fitting of down-quark mass ratios.  
Therefore, in the next section, we proposed a revised model 
for quark mass matrices keeping the model for the neutrino sectors.


\section{Phenomenology of quark mass matrices}

We present an explicit form of the quark mass matrices in our new model. 
In this paper, we put the following assumptions for a phenomenological 
forms of quark mass matrices $M_u$ ($M_u^{1/2}$) and $M_d$:

\noindent (i) Differently from the previous model \cite{O3PLB09}, 
we regard that
not only $\langle Y_u \rangle$ (also  $\langle \Phi_u \rangle$)
but also $\langle Y_d \rangle$ are real, i.e. $\alpha_d=0$ in 
Eq.(2.10).
Instead, we consider that $CP$ violation in the quark sector 
originates in a phase matrix $\langle P_d \rangle_e = v_{Pd} 
P_d \equiv v_{Pd} {\rm diag} (e^{i\phi_1}, e^{i\phi_2},e^{i\phi_3})$
 which does not affect the down-quark mass ratios, 
but does only the CKM mixing. 
Namely the quark mass matrices $M_u^{1/2}$ and $M_d$ is given by
$$
M_u^{1/2} \propto M_e^{1/2} ( {\bf 1} + a_u X) M_e^{1/2} 
+\xi_u \left( M_e^{1/2} M_e^{1/2} ( {\bf 1} + a_u X) 
+ ( {\bf 1} + a_u X) M_e^{1/2} M_e^{1/2} \right) 
+m_{0u} {\bf 1},
\eqno(3.1)
$$
$$
M_d \propto P_d \left[ M_e^{1/2} ( {\bf 1} + a_d X) M_e^{1/2} 
+\xi_d \left( M_e^{1/2} M_e^{1/2} ( {\bf 1} + a_d X) 
+ ( {\bf 1} + a_d X) M_e^{1/2} M_e^{1/2} \right) 
+m_{0d} {\bf 1}\right] P_d ,
\eqno(3.2)
$$
so that the CKM matrix $V$ is given by $V=R_u^T P_d R_d$, where
$R_u$ and $R_d$ are defined by $R_u^T M_u^{1/2} R_u =
{\rm diag}(+\sqrt{m_u}, -\sqrt{m_c}, +\sqrt{m_t})$ 
(for $a_u \sim -1.8$) and 
$R_d^T (P_d^\dagger M_dP_d^\dagger) R_d = {\rm diag}(m_d, m_s, m_b)$, 
respectively. 

\noindent (ii) Similarly to Eq.(2.15), we assume the 
$\xi_q$ terms which originate in the reordering of the fields
with the same quantum numbers.

\noindent (iii) Since only two of the three phase parameters 
$\phi_1$, $\phi_2$ and $\phi_3$ in the phase matrix 
$P_d ={\rm diag} (e^{i\phi_1}, e^{i\phi_2},e^{i\phi_3})$ are 
physically independent parameters. 
For convenience, we take $\phi_3=0$. 

\noindent (iv) It is better that the parameter number is 
as few as possible. 
We consider that the first term is dominant in Eq.(3.1)
[and also Eq.(3.2)], and we will consider $\xi_q$ and 
$m_{0q}$ terms as the need arises. As seen later, we can do fitting without $\xi_d$ and 
$m_{0u}$ terms.

Of course, we consider that these relations are derived from
SUSY vacuum conditions for a given superpotential $W$. 
However, prior to investigating the superpotential form, 
from a phenomenological point of view, 
we would like to investigate whether there is a possible parameter 
region or not in the present model.
A Yukawaon model for the phenomenological forms (3.1) and (3.2)
will be discussed in the next section.
 
Since the mass spectra $M_u^{1/2}(a_u)$ and $M_d(a_d)$ 
have the same behavior for the parameter $a_q$ ($a_u$ and
$a_d$), we illustrate the mass spectra versus $a_q$ 
in the limit of $\xi_q=0$ in Fig.2.
(The mass values in Fig.2 read $(\sqrt{|m_u|}, \sqrt{|m_c|},
\sqrt{|m_t|})$ and $(|m_d|, |m_s|, |m_b|)$ for the up- and 
down-quark sectors, respectively. 

\begin{figure}[t!]
  \includegraphics[width=78mm,clip]{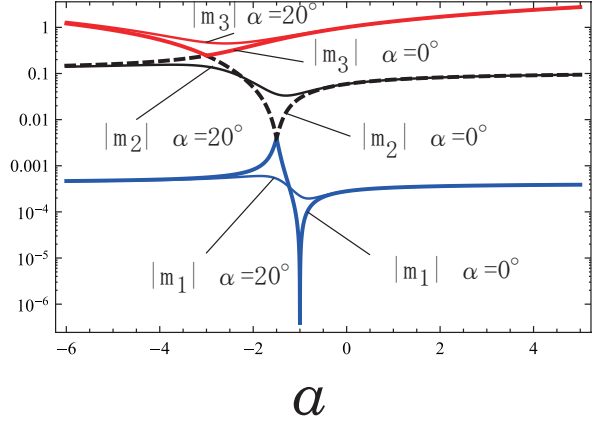}
  
  \caption{Eigenvalues $|m_1|,|m_2|,|m_3|$ versus a parameter
$a$ in a mass matrix $M= M_e^{1/2} ( {\bf 1} + a e^{i\alpha} X) 
M_e^{1/2}$ with $\alpha=0$ (thick curves). 
For reference, a case with $\alpha=20^\circ$ is also 
illustrated by thin curves in the figure.
$(|m_1|,|m_2|,|m_3|)$ read $(\sqrt{|m_u|}|, \sqrt{|m_c|}, 
\sqrt{|m_t|})$ and $(|m_d|, |m_s|, |m_b|)$ for 
up- and down-quark sectors, respectively.
Numerical values of the eigenvalues are given in a unit of 
$(m_e +m_\mu +m_\tau)$.    
}
  \label{masslevel}
\end{figure}

In the present model, too, the model for the 
up-quark sector and neutrino sector 
is essentially unchanged from the previous model \cite{O3PLB09}
except for the $\xi_u$ term given in (3.1).  
For reference, in Fig.3, we illustrate the up-quark mass ratios 
$\sqrt{m_u/m_c}$ and $\sqrt{m_c/m_t}$ versus $a_u$ and $\xi_u$.
As seen in Fig.3, there are two set of the solution $(a_u, \xi_u)$
[regions (i) and (ii) illustrated in Fig.3] 
which can give reasonable up-quark mass ratios.
However, the region (ii) cannot give reasonable CKM mixing
parameters. 
Hereafter, by taking fitting of neutrino mixing parameters 
into consideration, too, we will take $a_u=-1.764$ and 
$\xi_u=0.0070$ in the region (i).
The choice of $(a_u, \xi_u)=(-1.764,0.0070)$ can give
up-quark mass ratios
$$
\sqrt{\frac{m_u}{m_c} }=0.0619, \ \ \ \ 
\sqrt{\frac{m_c}{m_t} }=0.0559.
\eqno(3.3)
$$

\begin{figure}[t!]
  \includegraphics[width=78mm,clip]{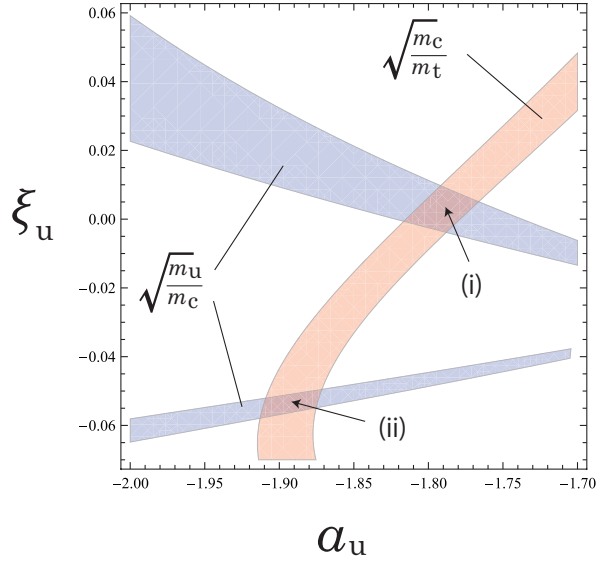}
  
  \caption{Allowed region in the $a_u$ - $\xi_u$ plane obtained from the up-quark mass ratios 
$\sqrt{m_u/m_c}$ and $\sqrt{m_c/m_t}$.
The shaded areas are consistent with the observed values given
in Eq.(2.13).
}
  \label{xiu-au}
\end{figure}

In model building of the down-quark sector, we give 
the down-quark mass ratio $m_s/m_b$ preference rather 
than $m_d/m_s$, because it is not so difficult to adjust 
the ratio $m_d/m_s$ without affecting the CKM parameter 
fitting as we demonstrate later.
In Fig.4, we illustrate behavior of $m_s/m_b$ versus $a_d$. 
As seen in Fig.4, there are three regions which can give
reasonable mass ratio $m_s/m_b$. 
However, the regions (ii) and (iii) cannot give reasonable
CKM mixing parameters.
(The region (ii) corresponds to a parameter region adopted
in the old model \cite{O3PLB09}.)
Hereafter, we will show the region (i) (i.e. $a_d\sim -17$) 
in detail.

\begin{figure}[t!]
  \includegraphics[width=78mm,clip]{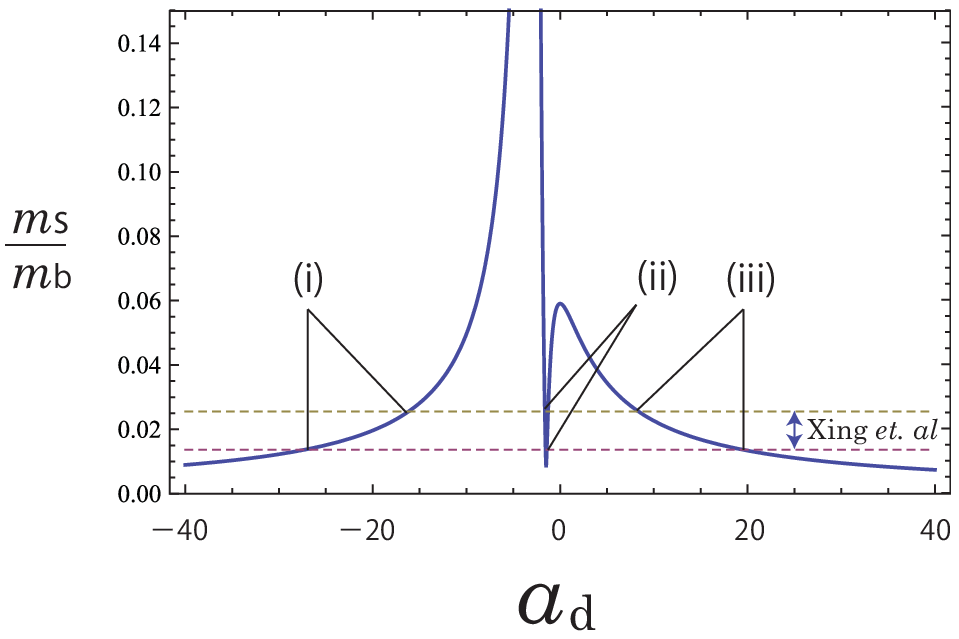}
  
  \caption{$a_d$ dependence of $m_s/m_b$. 
  The dotted lines show the observed down-quark
  mass ratio $m_s/m_d=0.019+0.006$ and $0.019-0.006$.
}
  \label{ckm-phi}
\end{figure}

Next, we investigate possible parameter regions 
which can give reasonable CKM mixing parameters.
We take $|V_{us}|$, $|V_{ub}|$, $|V_{cb}|$ and
$|V_{td}|$ as four independent parameters in 
the CKM matrix.
In Fig.5, we illustrate allowed regions in the $\phi_1$ - $\phi_2$ plane obtained from $|V_{ij}|$ with $|V_{ij}|_{obs}$ under $a_u=-1.764$, $\xi_u=0.0070$ 
and $a_d=-16.6$  whose values are obtained 
from global best fit.  
As seen in Fig.5, the value of $\phi_2 \simeq 180^\circ$ is in favor
of the observed CKM mixing parameters.
The case with $\phi_2=180^\circ$ is also 
illustrated in Fig.6. 
It is interesting that $|V_{ij}|$ take their minimum 
at $\phi_1 \simeq 180^\circ$.
From Fig.5 We find that $\phi_1 \simeq \pm 16^\circ +180^\circ$ is 
in favor of the observed CKM mixing parameters.

\begin{figure}[t!]
  \includegraphics[width=78mm,clip]{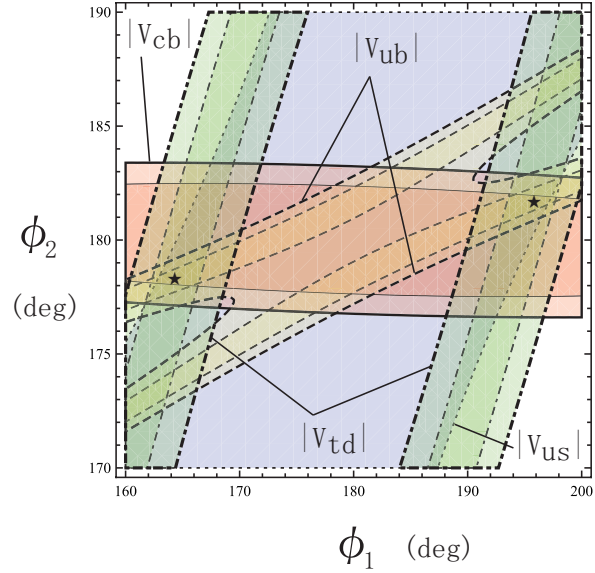}
  
  \caption{Allowed regions in the $\phi_1$ - $\phi_2$ plane 
obtained from the CKM mixing parameters $|V_{ij}|$. 
Shaded areas are consistent  with the observed values  $|V_{ij}|_{obs}$ in Eq.(2.18) .
The parameter values of $a_u$, $\xi_u$ and $a_d$ are
chosen as $a_u=-1.764$, $\xi_u=0.0070$ and $a_d=-16.6$
respectively. The star ($\star$) indicates the best fit 
points [see Eq.(3.4)].
}
  \label{fig5}
\end{figure}

\begin{figure}[t!]
  \includegraphics[width=78mm,clip]{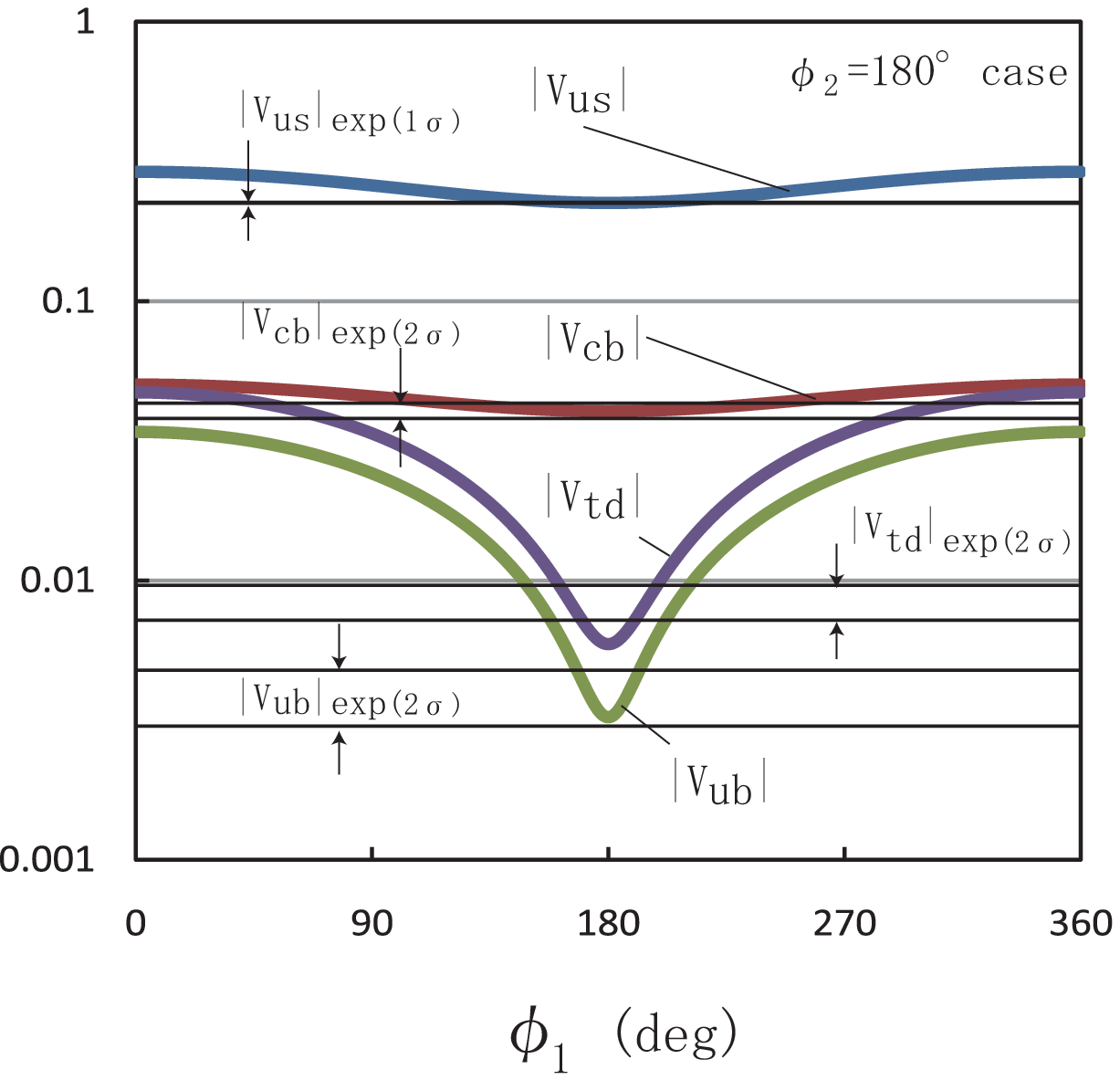}
  
  \caption{$\phi_1$ dependence of the CKM mixing parameters 
$|V_{ij}|$ in the case $\phi_2=180^\circ$. In the case of $\phi_2=180^\circ$, $\phi_1=191^\circ$ and $169^\circ$ are in favor of all the observed $|V_{ij}|_{exp}(2\sigma) $. 
}
  \label{fig6}
\end{figure}

In conclusion, our best hit parameters  are 
$$
a_u = -1.764, \ \ \ \xi_u=0.0070, \ \ \ a_d =-16.6,
\ \ \ \phi_1=196.0^\circ \ (164.0^\circ) , \ \ \ 
\phi_2= 181.5^\circ \ (178.5^\circ) 
\eqno(3.4)
$$
together with $\xi_d=0$, 
and then we obtain the predicted CKM mixing parameters
$$
|V_{us}|=0.2259, \ \ \ |V_{cb}|=0.04141, \ \ \ 
|V_{ub}|=0.00418, \ \ \ |V_{td}|=0.00854.
\eqno(3.5)
$$
However, the parameter value $a_d=-16.6$ gives 
considerably small value of $m_d/m_s$, i.e. 
$m_d/m_s=0.00358$. 
In order to correct this wrong value, we must 
take $m_{0d}$ with a non-zero value.
By taking a value 
$$
{m_{0d}}/m_0=-0.0061 ,
\eqno(3.6)
$$
where $m_0\equiv m_e+m_\mu+m_\tau$, 
we obtain the reasonable down-quark mass ratios
$$
\frac{m_d}{m_s}=0.0529, \ \ \ \ \frac{m_s}{m_b}= 0.0231 ,
\eqno(3.7)
$$
without affecting the CKM mixing parameters.

On the other hand, for neutrino mixing parameters,
the model is essentially the same as before.
In Fig.7, we illustrate $\xi_\nu$ dependence of 
the neutrino mixing parameters.
As seen in Fig.7, the model can predict 
$\sin^2 2\theta_{atm} \simeq 1$ and $|U_{13}|^2\simeq 0$
independently of $\xi_\nu$. 
The value of $\xi_\nu$ is determined from the observed value
$\tan^2 \theta_{solar}=0.457^{+0.038}_{-0.041}$.
The value of $\xi_\nu$ and the neutrino mixing parameters
are listed in Table I.

\begin{figure}[t!]
  \includegraphics[width=100mm,clip]{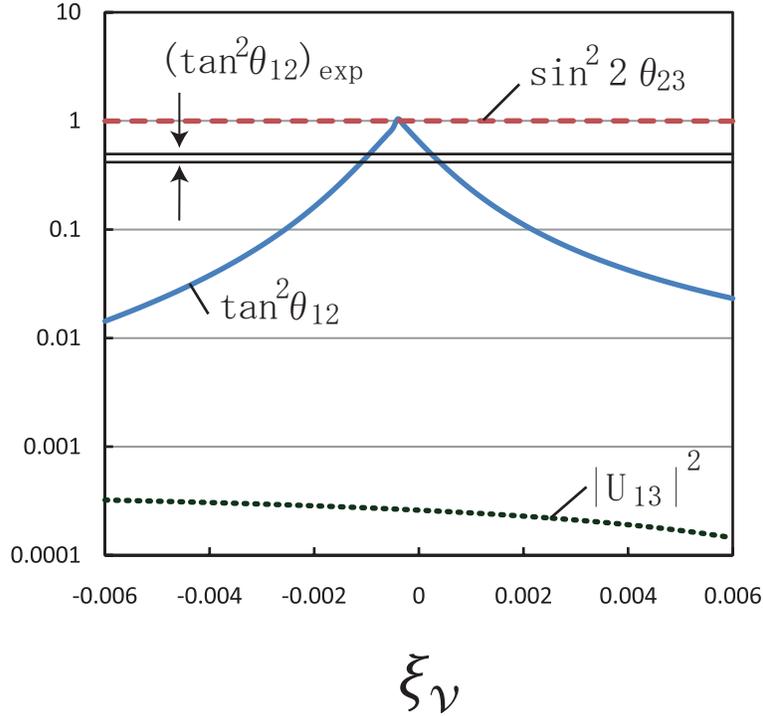}
  
  \caption{$\xi_\nu$ dependence of the neutrino mixing parameters 
$\tan^2\theta_{solar}$ (solid curve), $\sin^2 2\theta_{atm}$ 
(dashed curve) and $|U_{13}|^2$ (dotted curve).  
The up-quark mass matrix parameters are chosen
as $a_u=-1.764$ and $\xi_u=0.0070$.
}
  \label{fig7}
\end{figure}

\begin{table}
\begin{tabular}{cccc} \hline
$\xi_\nu$    &  $\tan^2 \theta_{solar}$ & $\sin^2 2\theta_{atm}$
& $|U_{13}|^2$ \\ \hline
$+0.00031$ & $ 0.457$ & $0.999$ & $ 2.56\times 10^{-4}$ \\
$-0.00102$ & $0.457$ & $0.998$ & $2.74 \times 10^{-4}$ \\
\hline
\end{tabular}

\caption{
Input value of $\xi_\nu$ and predicted values of the 
neutrino mixing parameters. The up-quark mass matrix
parameters are chosen as $a_u=-1.764$ and $\xi_u= 0.0070$ which can
give reasonable up-quark mass ratios.
}
\end{table}

Let us summarize above phenomenological considerations for 
the mass matrices for quarks and neutrinos. 
By taking the phenomenological considerations $\xi_d=0$ 
and $m_{0u}=0$ into consideration,  
we have adopted the quark mass matrices $M_u^{1/2}$ and $M_d$ given by
$$
M_u^{1/2} \propto M_e^{1/2} ( {\bf 1} + a_u X) M_e^{1/2} 
+\xi_u \left( M_e^{1/2} M_e^{1/2} ( {\bf 1} + a_u X) 
+ ( {\bf 1} + a_u X) M_e^{1/2} M_e^{1/2} \right) ,
\eqno(3.8)
$$
$$
M_d \propto P_d \left[ M_e^{1/2} ( {\bf 1} + a_d X) M_e^{1/2} 
+m_{0d} {\bf 1}\right] P_d .
\eqno(3.9)
$$
On the other hand, the neutrino mass matrix is given by Eqs.(2.14)--(2.16).
By using these mass matrices with the 7 free parameters, 
$a_u$, $\xi_u$, $a_d$, $\phi_1$, $\phi_2$, $m_{0d}$, and $\xi_{\nu}$, 
we have searched systematically for the parameter values 
which can give reasonable 4 quark mass ratios,  
4 CKM quark mixing parameters, and 3 neutrino mixing parameters. 
(Although the values of $(m_e, m_\mu, m_\tau)$ play an essential 
role in the present model, we have fixed those to the running 
mass values at $\mu=m_Z$, so that we do not count those as 
free parameters.)


\section{Superpotential}

In this section, by taking the phenomenological results 
with $\xi_d=0$ and $m_{0u}=0$ in the previous section
into consideration, we discuss a possible form of the
 superpotential $W$ assuming an O(3) family symmetry.
Since we consider the effective theory with $\Lambda \sim 10^{14}$
GeV, at present, it is not our chief concern whether O(3) is local
or global.
For the moment, we assume that O(3) is global. It should be noted that 
the massless states are harmless because $\Lambda$ takes an extreme large 
value $\Lambda \sim 10^{14}$ GeV 
\cite{masslessY}.
Under the O(3) family symmetry and conservations of U(1)$_X$ 
and $R$ charges given in Table II, we obtain the following
form of $W$:
$$
W=W_e +W_R + W_u +W_d ,
\eqno(4.1)
$$
$$
W_e= \mu_e {\rm Tr}[Y_e\Theta_e] + \lambda_e 
{\rm Tr}[\Phi_e\Phi_e\Theta_e] ,
\eqno(4.2)
$$
$$
W_R = \mu_R {\rm Tr}[Y_R \Theta_R]
+ \frac{\lambda_R}{\Lambda} \left\{
{\rm Tr}[(Y_e P_u \Phi_u + \Phi_u P_u Y_e)\Theta_R]
+
\xi_\nu {\rm Tr}[(P_u Y_e  \Phi_u 
+ \Phi_u  Y_e P_u)\Theta_R] \right\},
\eqno(4.3)
$$
$$
W_u = \mu_u {\rm Tr}[Y_u\Theta_u] + \lambda_u 
{\rm Tr}[\Phi_u\Phi_u \Theta_u]
$$
$$
+ \mu'_u {\rm Tr}[\Phi_u\Theta'_u] + 
\frac{\lambda'_u}{\Lambda} \left\{ 
{\rm Tr}[\Phi_e S_u \Phi_e \Theta'_u] 
+\xi_u {\rm Tr}[(\Phi_e \Phi_e S_u +S_u \Phi_e \Phi_e) 
\Theta'_u] \right\} ,
\eqno(4.4)
$$
$$
W_d = \frac{\lambda_d}{\Lambda} 
{\rm Tr}[P_d Y_d P_d \Theta_d]  
+ \frac{\lambda'_d}{\Lambda} 
{\rm Tr}[\Phi_e S_d \Phi_e \Theta_d] 
 +\mu_{0d} {\rm Tr}[E_{0d} \Theta_d] ,
\eqno(4.5)
$$
where, for convenience, we have denoted linear 
combinations of fields $E_q$ and $X_q$ as
$S_q=E_q + a_q X_q$ ($q=u,d$) and 
$\langle E_q \rangle =v_{Eq} {\bf 1}$ and
$\langle X_q \rangle =v_{Xq} X$.

Among the SUSY vacuums which are derived from the
superpotential (4.1), we take only a vacuum
with $\langle \Theta_e \rangle =\langle \Theta_R \rangle
=\langle \Theta_u \rangle =\langle \Theta'_u \rangle
=\langle \Theta_d \rangle =0$.
Therefore, we can obtain VEV relations 
(2.6), (2.15), (3.1) and (3.2) from SUSY vacuum conditions 
$\partial W/\partial \Theta_e =0$, 
$\partial W/\partial \Theta_R =0$,
$\partial W/\partial \Theta'_u =0$ and 
$\partial W/\partial \Theta_d =0$, respectively. 
Since other conditions, for example, 
$\partial W/\partial Y_e =0$, and so on,
inevitably contain a field $\Theta_A$ ($A=u,d,\cdots$),
they cannot play effective roles in the VEV relations.
(Although we did not give an explicit form of $W(\Phi_e)$
in the present paper, we assume that $W(\Phi_e)$ also
contains $\Theta$ fields.
For the form of $\langle \Phi_e \rangle$,
Eq.(2.7), we will use the charged lepton mass values
at $\mu=m_Z$. )
One of merits to introduce such  $\Theta$ fields is 
that we do not need to consider contributions from 
higher dimensional terms with the form 
$({\rm Tr}[\cdots \Theta_A])^n$  ($n\geq 2$), 
because $\partial W/\partial \Theta_A$ from such 
a higher dimensional term always contains
$\Theta_A$ more than one, so that such a term becomes
vanishing.

Let us emphasize a role of the $R$ charges:
By assuming the $R$ charge conservation with the $R$ charge 
assignment given in Table II, we can forbid all of higher dimensional 
terms with $(1/\Lambda)^n$ ($n \geq 2$) except for the terms given by 
Eqs.(4.2) - (4.5).
(However, for this purpose, we must assume that 
our K\"{a}hler potential is given by a canonical 
(minimal) form.)
We also note that if we assume U(1)$_X$ only, 
the assignments can allow unwelcome terms in the superpotential, 
for example, ${\rm Tr}[S_u P_u]$, 
${\rm Tr}[\Phi_u P_u \Theta_e]$, and so on.
Such terms can be forbidden by assuming suitable $R$ 
charge assignments.
For example, when we take $R$-charges as
$R(\ell)=1-r$, $R(e^c)=R(\nu^c)=R(q)=R(u^c)=R(d^c)=1$
and $R(H_u)=R(H_d)=0$, we can forbid the terms
${\rm Tr}[S_u P_u]$ and  
${\rm Tr}[\Phi_u P_u \Theta_e]$ by taking $R$ charges
of other fields as given in Table II.

\begin{table}
\begin{tabular}{ccccccccccccccccc} \hline
Fields & $Y_e$ & $\Phi_e$ & $\Theta_e$ & $Y_R$ 
& $\Theta_R$ & $Y_u$ & $\Phi_u$ & $P_u$ & $\Theta_u$ 
& $\Theta'_u$ & $E_u, X_u$ & $Y_d$ & $P_d$ & $\Theta_d$ 
& $E_d,X_d$ & $E_{0d}$ \\
$Q_X$  & $x_e$ & $\frac{1}{2} x_e$ & $-x_e$ & 
$2 x_e$  & $-2x_e$ & $x_u$ & $\frac{1}{2} x_u$ & 
$x_e-\frac{1}{2} x_u$& $-x_u$ & $-\frac{1}{2} x_u$ 
& $\frac{1}{2} x_u-x_e$ & $x_d$ & $x_P$ & 
$-(x_d+2 x_P)$ & $x_d +2 x_P -x_e$ & $x_d + 2x_P$ \\ 
$R$ charge & $r$ & $\frac{1}{2}r$ & $2-r$ & $0$ & $2$ &
$0$ & $0$ & $0$ & $2$ & $2$ & $-r$ & $0$ & $0$ & 
$2-r$ & $0$ & $r$ \\
\hline
\end{tabular}

\caption{ U(1)$_X$ charges of the Yukawaons. 
For the time being, we assign different charges for 
the fields $E_u$ and $E_d$ ($X_u$ and $X_d$) 
by assuming that those are different fields. 
For $R$ charges, see text.
}
\end{table}

In the phenomenological study in Sec.III, the VEV values
of $\Phi_e$ play an essential role in evaluating 
the predicted values. 
Although it has been tried to build a model 
\cite{Sumino09JHEP,e-spec_10PLB} which gives VEV 
spectrum (2.7), it is not clear whether such a model can 
be applicable or not to the present model straightforwardly.
In this paper, we do not give a superpotential form which
can lead to the VEV spectrum (2.7).
We have just assumed the VEV value given by Eq.(2.7), where 
we have used the values of charged lepton masses at 
the scale $\mu=m_Z$.

Also, so far, we have not given superpotential forms which 
lead to
VEV matrices $\langle E_q \rangle$, $\langle X_q \rangle$, 
$\langle P_u \rangle$ and $\langle P_d \rangle$. 
In general, any Hermitian VEV matrix $\langle A \rangle$ 
can be obtained from a superpotential
$$
W = \lambda_1 ({\rm Tr}[A])^3 +\lambda_2 {\rm Tr}[AA]
{\rm Tr}[A] + \lambda_3 {\rm Tr}[AAA] .
\eqno(4.6)
$$
[However, we must assign a U(1)$_X$ charge
$-3Q_X(A)$ 
for the coefficients $\lambda_i$ ($i=1,2,3$).]
For example, $(\lambda_1, \lambda_2,\lambda_3)=
(1/6,-1/2,1/3)$ (i.e. $W={\rm det}A$) gives 
the form $\langle X\rangle$ given in Eq.(2.11).
However, this method is not applicable to the form 
$\langle P_d \rangle$, because the VEV matrix is not
Hermitian.
In this paper, we have assumed these ad hoc VEV forms.


\section{Concluding remarks}

In conclusion, we have proposed a phenomenological quark 
and lepton mass matrices based on a Yukawaon model.
In Sec.II, we have demonstrated that the previous 
model \cite{O3PLB09}, in which the $CP$ violation originates 
only in the complex parameter $a_d$, cannot give reasonable 
CKM mixing values even if at the cost of the quark mass ratios. 
Differently from the previous model, in the present model, 
$CP$ violating phases are introduced  
in the phase matrix $P_d$ given in Eq.(3.2). 
In the up-quark sector, we have considered the $\xi_u$ term 
in Eq.(3.1). This comes from the fact that the terms with  
another order of the fields, $\Phi_e\Phi_e S_u+S_u \Phi_e
\Phi_e$, cannot be, in general, forbidden compared 
with the order of $\Phi_e S_u \Phi_e$ because of the
same U(1)$_X$ charges.
A similar situation have been assumed in the
neutrino sector, too, i.e. the $\xi_\nu$ term 
in Eq.(4.3). 
(The values of $\xi_u$ and $\xi_\nu$ are very small.) 
In contrast to those sectors, 
in the down-quark sector, we have not considered such a 
$\xi_d$ term as well as an additional term
$P_d P_d Y_d +Y_d P_d P_d$ corresponding to $P_d Y_d P_d$ 
in Eq.(4.5). 
This is a result from the phenomenological study, and   
the theoretical reason for the absence is unknown 
at present.
Also we note that the phenomenological fit requires
the $m_{0d}$ term added to Eq.(3.2), but it does not 
need an $m_{0u}$ in Eq.(3.1).

Our numerical conclusions from the present systematical
study is summarized in Figs.~2-7.
Especially, as seen in Fig.7, the results 
$\sin^2 2 \theta_{atm} \simeq 1$ and $|U_{13}|^2 \le 0.005$
are insensitive to the value of the parameter $\xi_\nu$.
In other words, if $|U_{13}|^2 \sim 0.01$ (the possibility
was pointed out by Fogli, {\it et al.} \cite{Fogli}) 
is established experimentally, the present model will 
be ruled out, or it will need a drastic revision. 

We have been able to obtain reasonable 
parameter fitting not only for the observed lepton 
mixing but also for the observed quark mixing. 
However, the model still includes ad hoc assumptions. 
We consider that it is important to clarify what 
parts are problems to get a good fitting of the data 
for the next step of the investigation.
Our model building will proceed step by step.

\vspace{10mm}
\centerline{\large\bf Acknowledgments} 

The authors would like to thank T.~Yamashita for
his valuable and helpful comments, especially
on the effective theory.
One of authors (Y.K.) is supported by the Grant-in-Aid for
Scientific Research (C), JSPS, No.21540266.


\end{document}